\documentclass[prl,amsfonts,amssymb,floats,twocolumn,superscriptaddress,aps]{revtex4}

\usepackage[final,dvips]{epsfig}
\usepackage{graphicx}

\newcommand{\al}{\alpha}
\newcommand{\w}{\omega}

\begin{document}

\title[]
{
Equilibrium and non-equilibrium dynamics
of the sub-ohmic spin-boson model
}
\author{Frithjof B. Anders}
\affiliation{Fachbereich Physik, Universit\"at Bremen,
28334 Bremen, Germany}
\author{Ralf Bulla}
\affiliation{\mbox{Theoretische Physik III, Elektronische Korrelationen und
Magnetismus, Universit\"at Augsburg,
86135 Augsburg, Germany}}
\author{Matthias Vojta}
\affiliation{Institut f\"ur Theoretische Physik,
Universit\"at zu K\"oln, Z\"ulpicher Str. 77, 50937 K\"oln, Germany}
\date{April 2, 2007}

\begin{abstract}
Employing the non-perturbative numerical renormalization group method,
we study the dynamics of the spin-boson model,
which describes a two-level system coupled to a bosonic bath with
spectral density $J(\omega)\propto\omega^s$.
We show that, in contrast to the case of ohmic damping,
the delocalized phase of the sub-ohmic model cannot be characterized
by a single energy scale only, due to the presence of a non-trivial quantum phase
transition.
In the strongly sub-ohmic regime, $s\ll 1$, weakly damped coherent oscillations on short time
scales are possible even in the {\em localized} phase -- this is of crucial relevance,
e.g., for qubits subject to electromagnetic noise.
\end{abstract}


\maketitle


Models of quantum dissipation \cite{Leggett, Weiss} have gained significant attention
over the last years, due to their wide range of applications
from the effect of friction on the electron
transfer in biomolecules \cite{Garg} to the description
of the quantum entanglement between a qubit and its
environment \cite{dima,wilhelm} (for further applications
see Refs. \onlinecite{Leggett, Weiss}).

The simplest quantum-dissipative models belong to the class of impurity models
and consist of a small quantum system coupled to a bath of harmonic oscillators.
One familiar bosonic impurity model is the spin-boson model,
\begin{equation}
H=-\frac{\Delta}{2}\sigma_{x}+\frac{\epsilon}{2}\sigma_{z}+
\sum_{i} \omega_{i}
     a_{i}^{\dagger} a_{i}
+\frac{\sigma_{z}}{2} \sum_{i}
    \lambda_{i}( a_{i} + a_{i}^{\dagger} ).
\label{eq:sbm}
\end{equation}
It describes a generic two-level system, represented by the Pauli matrices $\sigma_j$,
which is linearly coupled to a bath of harmonic oscillators,
with creation operators $a_i^\dagger$ and frequencies $\omega_i$.
The bare tunneling amplitude between the two spin states
$|\uparrow\rangle$, $|\downarrow\rangle$ is given by $\Delta$,
and $\epsilon$ is an additional bias (which is zero in the following,
except for the preparation of the initial state as discussed below).
The coupling between spin and bosonic bath is specified
by the bath spectral function
$
J(\omega)= \pi \sum_{i} \lambda_{i}^{2} \delta(\omega -\omega_{i})
$.
The asymptotic low-temperature behavior is determined by the low-energy
part of the spectrum.
Discarding high-energy details, the standard parametrization is
\begin{equation}
  J(\omega) = 2\pi\, \alpha\, \omega_c^{1-s} \, \omega^s\,,~ 0<\omega<\omega_c\,,\ \ \ s>-1
\label{power}
\end{equation}
where the dimensionless parameter $\alpha$ characterizes the
dissipation strength, and $\omega_c$ is a cutoff energy.

In case of ohmic dissipation, $s\!=\!1$,
a quantum transition of Kosterlitz-Thouless type
separates a localized phase at $\alpha \geq \alpha_c$,
displaying a doubly degenerate ground state,
from a delocalized phase at $\alpha<\alpha_c$ with a unique ground
state \cite{Leggett,Weiss}.
The delocalized regime is characterized by a finite effective tunnel
splitting, $\Delta_r$, between the two levels,
whereas the tunnel splitting renormalizes to zero in the localized phase.
For $\Delta \ll \omega_c$ the transition occurs at $\alpha_c=1$.

\begin{figure}[!t]
\epsfxsize=2.9in
\centerline{\epsffile{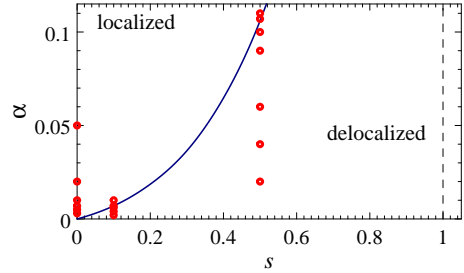}}
\caption{
Zero-temperature phase diagram of the spin-boson model from NRG \cite{nrgpara},
for fixed $\Delta=0.1$ and $\w_c=1$.
The axes denote the bath exponent $s$ and the dissipation strength $\al$.
The circles indicate the parameter values for which
results will be presented in this paper.
}
\label{fig:phd}
\end{figure}

The sub-ohmic case, $s<1$, \cite{dima,KM,spohn,BTV,turlakov} turns out to be different.
For $\Delta/\omega_c\to 0$ the system is localized for any non-zero
coupling \cite{Leggett,Weiss}, but
for large $\Delta$ a delocalized phase was argued to exist \cite{spohn,KM}.
We have recently \cite{BTV} shown, using an extension of the non-perturbative
Numerical Renormalization Group (NRG), that a {\em continuous} quantum phase transition
(QPT)
occurs for {\em all} $0<s<1$, in contrast to earlier proposals \cite{KM}.
The numerically determined equilibrium phase diagram is shown in
Fig.~\ref{fig:phd}.

Detailed studies of non-equilibrium properties, like the quantity
$P(t)=\langle \sigma_z(t)\rangle$ for a spin initially prepared in the $|\uparrow\rangle$
state, have mainly covered the ohmic damping case for small $\Delta$
\cite{Leggett,Costi96,KeilSchoeller2001}.
Here, weakly damped oscillations can only be observed for $\alpha<1/2$,
whereas $1/2<\alpha<1$ leads to overdamped behavior, i.e., exponential decay
of $P(t)$ to zero.
Finally, in the localized phase, $\al>1$, $P(t)$ decays to a finite value.
Studies of the dynamics in the sub-ohmic case
have so far been restricted to the use of perturbative methods \cite{shnirman}.
Considering that these miss the QPT and the localized phase
present for $s<1$, the validity of the corresponding results is questionable.

The purpose of this paper is to provide
non-perturbative results for the dynamics of the sub-ohmic spin-boson
model, for the whole range of model parameters.
Our main results can be summarized as follows:
Whereas the delocalized phase of the ohmic model
is dominated by a single energy scale, $\Delta_r$, only,
the sub-ohmic model is more complicated due to the presence of a second-order
QPT.
Near the transition a quantum critical (QC) crossover scale $T^\ast$ appears \cite{BTV},
which -- in the strongly sub-ohmic regime -- co-exists with a larger scale $\Delta_r$
that can still be identified with the renormalized tunnel splitting.
As we demonstrate below, $T^\ast$ vanishes at the transition
to the localized phase, whereas $\Delta_r$ stays finite.
The latter fact is crucial for the non-equilibrium dynamics:
we find the presence of coherent weakly damped oscillations, with frequency $\Delta_r$,
even in the localized phase, $\alpha > \alpha_c$, for $s\ll1$.
This is particularly relevant for the case $s=0$, related to so-called $1/f$ noise
in an electromagnetic environment.
Here, the equilibrium spin-boson model is always in the localized phase, however,
coherent oscillations are still possible for small $\alpha$ and short times.



{\it NRG.}
We employ Wilson's NRG method \cite{wilson} to study the sub-ohmic
spin-boson model, utilizing two recently developed extensions:
(i) Refs.~\onlinecite{BTV,BLTV} generalized the NRG at thermodynamic
equilibrium to impurity models with a {\em bosonic} bath.
(ii) Ref.~\onlinecite{neqnrg} proposed an algorithm to study
non-equilibrium dynamics in real time and applied it to the Anderson
and Kondo models.
(For an earlier non-equilibrium NRG approach see Ref.~\onlinecite{Costi97}.)
The time-dependent NRG  provides a spectral representation of
the time-independent Hamiltonian $H^f$, governing the time evolution for $t>0$,
at all energy scales using a complete basis set, and expresses the
real-time dynamics of observables in terms of a summation over reduced
density matrices~\cite{neqnrg}.
The latter contain all information on decoherence and dissipation.
Two independent bosonic NRG runs~\cite{BLTV} are required,
one for the initial density matrix and the other for the
approximate eigenbasis of  $H^f$.
In order to accurately simulate the continuum limit with a NRG chain of finite length,
we average over $N_Z$ different bath discretizations for a fixed
discretization parameter $\Lambda$.
All numerical results below are for temperature $T=0$, unless otherwise noted.


{\it Renormalization group flow and crossovers.}
To set the stage, we summarize the renormalization group flow of the spin-boson
model \cite{BTV,BLTV,VTB}.
In the following, the terms localized (delocalized)
are defined through the impurity entropy \cite{BLTV,entropyfoot}
being $\ln 2$ (zero).
Note that this has to be contrasted with localization in the sense
that a system initially prepared with the impurity spin in
one specified direction remains in this spin state
under time evolution. For any finite temperature, thermal
excitations destroy localization in this sense
(see Ref.~\onlinecite{Leggett}). For a discussion of this point,
in particular the connection between NRG flow and thermodynamic
properties, see Secs.~III and IV in Ref.~\onlinecite{BLTV}.

\begin{figure}[!t]
\epsfxsize=3.5in
\centerline{\epsffile{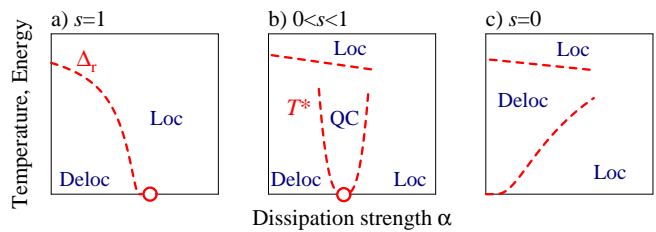}}
\caption{
Schematic crossover diagrams for the spin-boson model for different bath exponents $s$,
deduced from NRG, with regimes of localized, delocalized, and
quantum critical (QC) behavior.
The open dots denote QPTs,
for details see text.
(Note that the $\al$ scale is different for the three panels.)
}
\label{fig:xover}
\end{figure}

In the ohmic case, the flow for $\al<\al_c$ is from the localized towards the
delocalized phase (upon lowering the energy),
with $\Delta_r$ being the crossover energy scale.
Thus the behavior for energies or temperatures below $\Delta_r$ is
delocalized, Fig.~\ref{fig:xover}a.

For $0<s<1$ the critical fixed point controls the physics of the
QC region.
It is bounded by crossover lines $T^\ast \propto |\al-\al_c|^\nu$
and covers a large portion of the phase diagram due to the large value of the
correlation length exponent $\nu$, Fig.~\ref{fig:xover}b.
(In fact, $\nu$ diverges both as $s\to 1^-$ and $s\to 0^+$ \cite{VTB}.)
The critical fixed point merges with the localized (delocalized) one as
$s\to 1^-$ ($s\to 0^+$), implying that the characteristics of the
QC regime becomes more ``delocalized'' as $s$ is decreased
towards zero.
For $s$ not too close to 1 there also is a distinct crossover from the
high-temperature localized regime ($T\gg\Delta$) to the delocalized or
critical regimes.

Consequently, for $s=0$ and small $\al$ the flow is first from localized to
delocalized, with a crossover scale $\Delta_r$, finally to the localized fixed point
(which controls the ground state for any $\al$) -- the latter crossover is
characterized by the scale $T^\ast$, with $T^\ast \sim \omega_c \exp(-\Delta/\alpha\omega_c)$.
As shown in Fig.~\ref{fig:xover}c, the low-energy crossover at $s=0$ is
thus opposite to the one at $s=1$!

In general, we may expect coherent, weakly damped dynamics, with a rate $\Delta_r$,
in the delocalized regime (and in the QC one for small $s$ as well).


{\it Equilibrium dynamics.}
Let us focus on the Fourier transform $C(\w)$ of
the symmetrized equilibrium correlation function
$C(t) = \frac{1}{2}\langle [\sigma_z(t),\sigma_z]_+ \rangle$.
We start by noting the low-frequency asymptotics:
In the delocalized phase $C(\w) \propto \w^s$ \cite{KM,spzw},
whereas in the localized phase $C(\w)$ displays a $A \delta(\omega)$
contribution, which reflects the doubly degenerate ground state, and
$A$ plays the role of an order parameter
(equivalent to the thermodynamic expectation value $\langle \sigma_z \rangle$).
Turning to the finite-frequency behavior,
our numerics for $s=1$ reproduces the well-known results
(see Sec.~V in Ref.~\onlinecite{BLTV}), i.e., $C(\w)$
is dominated by a single peak at $\Delta_r$ which shifts to lower
frequencies with increasing $\al$ and disappears at the transition, $\al=\al_c$.
(Results for $s$ down to 0.8 are qualitatively similar.)

\begin{figure}[!t]
\epsfxsize=3.5in
\centerline{\epsffile{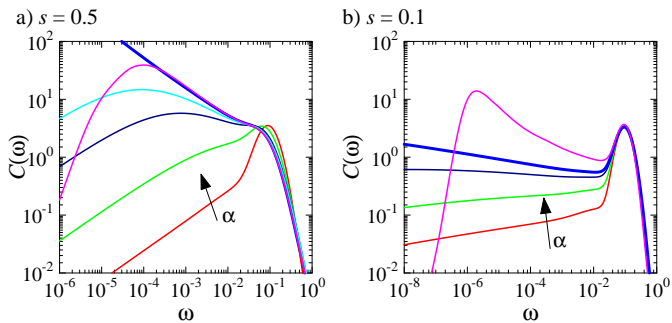}}
\caption{
Equilibrium spin correlation function $C(\w)$ for
a) $s=0.5$ and $\alpha = 0.02, 0.06, 0.09, 0.10, 0.107, 0.11$;
b) $s=0.1$ and $\alpha = 0.002, 0.004, 0.006, 0.007, 0.01$;
the thick curves are very close to the critical point
(for $s=0.5$ we find $\alpha_c=0.1065$ and $s=0.1$, $\alpha_c=0.0071$).
The curves correspond to the parameter values shown
as circles in Fig.~\ref{fig:phd}.
(The peak width at small $\alpha$ is an artifact of the NRG broadening.)
}
\label{fig:c05}
\end{figure}

For smaller $s$ the influence of the critical fixed point
becomes increasingly visible, see Fig.~\ref{fig:c05}a for $s=0.5$.
For small $\al$ we observe that $C(\w)$ is dominated by a peak close
to $\Delta$, as expected.
However, with increasing $\al$ weight is transferred to smaller frequencies,
leaving a shoulder feature close to $\Delta$ intact.
For $\al \lesssim \al_c$, $C(\w)$ has a pronounced peak at $\w\sim T^\ast$ --
this peak separates the QC divergence $C(\w) \propto \w^{-s}$
at intermediate energies from the low-energy $\w^s$ behavior.
At criticality, $T^\ast\to 0$, and $C(\w) \propto \w^{-s}$ down to lowest energies.

\begin{figure}[!b]
\epsfxsize=3.5in
\centerline{\epsffile{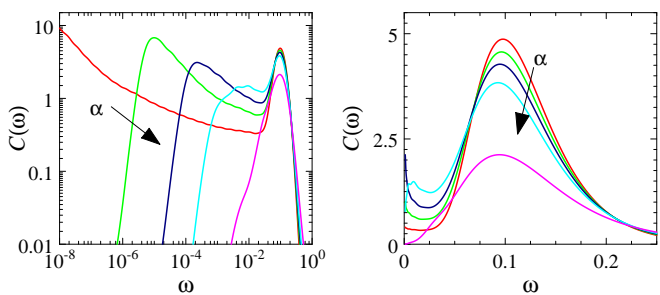}}
\caption{
Equilibrium spin correlation function $C(\w)$ for $s=0$
and $\alpha = 0.003, 0.005, 0.007, 0.01, 0.02$.
Both panels show the same data, but the right one has a linear $\w$ scale.
The $A\delta(\w)$ contribution is not shown.
}
\label{fig:c0}
\end{figure}

A further decrease of the bath exponent $s$ -- see Fig.~\ref{fig:c05}b for $s=0.1$ --
shows that the
high-frequency peak in the vicinity of $\Delta$ is suppressed more slowly
upon increasing $\al$, reflecting the fact that the QC behavior
is more ``delocalized'' for small $s$.
In particular, this peak, which is usually taken as indication for the presence
of coherent weakly damped dynamics, survives even for $\al>\al_c$ (!).

Fig.~\ref{fig:c0} for $s=0$ illustrates this fact:
here the ground state is localized for any $s$, but a well-defined peak
at $\Delta_r$ is visible for all $\al < 0.05$ (for $\Delta/\omega_c=0.1$).
Further, there is a low-energy peak at $T^\ast$ for small $\al$,
in addition to the $A \delta(\omega)$ contribution (not plotted)
characteristic of the localized phase.

Parenthetically, we note that the low-energy part of $C(\w,T)$ near $\al=\al_c$
is expected to show universal scaling behavior, including $\w/T$ scaling,
as the critical fixed point obeys hyperscaling properties for all $0<s<1$ \cite{VTB}.


{\it Non-equilibrium dynamics.}
We now present results for $P(t)$, obtained with the non-equilibrium
generalization of the NRG~\cite{neqnrg}.
The spin is prepared in the $|\!\!\uparrow\rangle$ state, by calculating
the equilibrium density matrix of the initial Hamiltonian $H^i$ with a large
spin polarization energy $\epsilon/\omega_c=100$ and $\Delta=0$.
At $t=0$, $\epsilon$ is switched off, and $\Delta$ is set to
$\Delta/\omega_c=0.1$. This defines $H^f$, governing the time evolution
of the initial density operator.

\begin{figure}[!t]
\centering\includegraphics[width=3.2in]{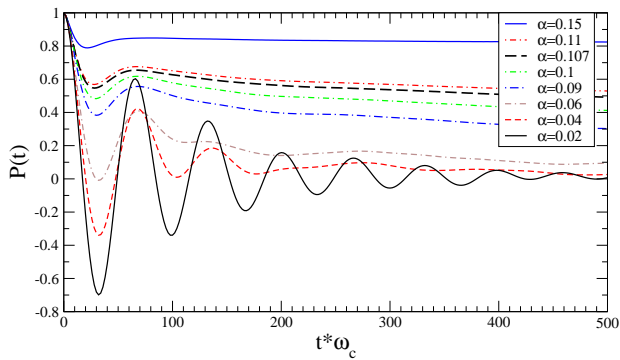}
\caption{
Time evolution of $P(t)$ for $s=0.5$, $\Delta/\omega_c=0.1$, and
various values of $\alpha$ ($\alpha_c=0.1065$).
}
\label{fig:p05}
\end{figure}

Fig.~\ref{fig:p05} shows $P(t)$
for $s=0.5$, $\Delta/\omega_c=0.1$, and various values of
$\alpha$. For weak coupling, i.e.\ $\alpha\ll \alpha_c$, a damped oscillatory
behavior is found. Increasing $\alpha$ suppresses the oscillations
at longer time scales but maintains the initial ones.
A very shallow oscillation at short times can be
seen even for $\alpha>\alpha_c$ which is consistent with the $C(\w)$ data
presented in Fig.\ \ref{fig:c05}.
For $\alpha>\alpha_c$ $P(t)$ shows only a very weak time dependence for intermediate
times and is expected to approach a finite value $P(\infty)$ for $t\to \infty$,
an indication of localization (see below).

Let us now turn to $s=0$, Fig.~\ref{fig:p0}.
Even though any finite value of $\alpha$ places the model in the localized phase
in equilibrium, the non-equilibrium dynamics of $P(t)$ clearly exhibits oscillatory
behavior for small $\alpha$.
Roughly speaking, the time evolution mimics the RG flow and thus corresponds
to a reduction of temperature in the phase diagram Fig.~\ref{fig:xover}c:
At short times the physics is governed by the delocalized fixed
point; it crosses over to the localized fixed point at long time
scales.
Upon increasing $\alpha$ the time scale of the crossover from oscillatory to damped
behavior in $P(t)$ is reduced, consistent with the low-temperature crossover
line in Fig.~\ref{fig:xover}c.

For small damping, the decay of the $P(t)$ oscillations is exponential.
Comparing the decay rate with the popular Bloch-Redfield approach \cite{redfield}
gives deviations of less than $10\%$ for $s=0.1$, $\alpha=0.002$,
but the agreement becomes worse for larger $\alpha$ or smaller $s$.
For all $s<1$ we observe that with increasing $\alpha$ the frequency of the
initial oscillations in $P(t)$ first decreases
(consistent with the result of straightforward perturbation theory),
but for larger $\alpha$ and small $s$ then slightly increases with $\alpha$,
indicating that one leaves the perturbatively accessible regime.
[$C(\w)$ in Figs.~\ref{fig:c05}, \ref{fig:c0} is inconclusive w.r.t. the peak position
due to the logarithmic broadening with the NRG.]

The long-time limit $P(\infty)$ measures the ``degree of localization'',
equivalent to the order parameter $\langle \sigma_z \rangle$ in equilibrium
[and proportional to the $A$-coefficient of the $\delta(\w)$ contribution to
$C(\w)$].
$P(\infty)$ vanishes continuously at the transition for any $0\le s<1$.
[For the smallest value of $\alpha$ shown in Fig.~\ref{fig:p0} for $s=0$ ($\alpha=0.003$),
the $P(t)$ oscillates around $P(\infty)\approx 0.1$.]
Note the difference to the ohmic case, $s=1$,
where $P(\infty)$ jumps from a finite value for $\alpha>\alpha_c$ to zero
$\alpha<\alpha_c$.

A few remarks are in order.
On general grounds, one may expect a power-law decay in the long-time limit
of $P(t)$ at the critical point, $\al=\al_c$, for all $0<s<1$.
This power law will be cutoff away from criticality, on a time scale $\propto 1/T^\ast$.
At present, the accuracy of the numerical data in non-equilibrium is not
sufficient to verify this.
Since the short-time behavior of $P(t)$ is dominated by the high-energy
properties, i.e., by $\Delta_r$, small but finite temperatures do not alter the
response on time scales shown in Figs.\ \ref{fig:p05} and
\ref{fig:p0}. At finite $T$ we only observe a decay of $P(t)$ to zero
for long times and parameters corresponding to the localized
phase (not shown).


\begin{figure}[!b]
\centering\includegraphics[width=3.2in]{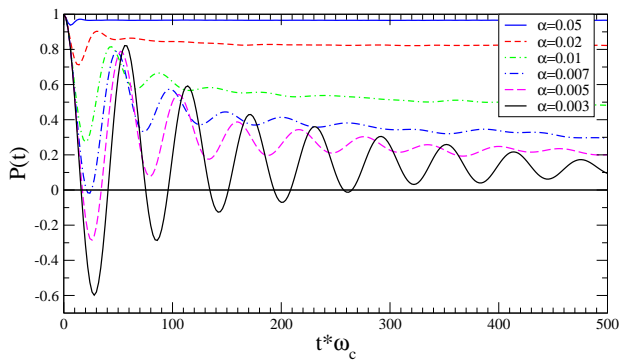}
\caption{
Time evolution of $P(t)$ for $s=0$, $\Delta/\omega_c=0.1$, and various values of
$\alpha$ ($\alpha_c=0$ here).
}
\label{fig:p0}
\end{figure}


{\it Summary.}
We have studied the dynamics of the sub-ohmic spin-boson model,
both in equilibrium and non-equilibrium, using non-perturbative
numerical renormalization group techniques.
The model displays a continuous QPT for
all bath exponents $0<s<1$, and this leads to highly non-trivial dynamical
properties.
In contrast to the ohmic situation, the sub-ohmic case cannot be characterized
by a single energy scale only. This is particularly striking for $s\ll 1$:
while the low-energy, long-time behavior is dominated by the
presence of nearly critical fluctuations over a large regime of parameters,
the behavior at short times or elevated energies derives from
weakly renormalized tunneling, which can be coherent even in situations
with a localized thermodynamic ground state.

Our results imply that perturbative methods, only capturing the
renormalized coherent tunneling, may be applied to the sub-ohmic model
at short times, but clearly fail in the long-time limit.
Furthermore, approximations aiming on a description of the spin-boson physics
in terms of a single energy scale $\Delta_r$ only, as done in Refs.~\onlinecite{KM,turlakov},
are not applicable in the strongly sub-ohmic regime.
Last not least, let us emphasize that the sub-ohmic situations of $s\!=\!1/2$ and $s\!=\!0$
are of immediate experimental relevance, e.g. for the description of electromagnetic
transmission lines \cite{devoret1} and general $1/f$ noise, respectively.
Furthermore, $s=1/2$ baths may be realized in the context of effective impurities
in ultracold gases \cite{sbgases}.

We thank S. Kehrein, S. Kohler,
 A. Rosch, S. Tornow, and W. Zwerger for discussions,
and A. Schiller and N.-H. Tong for collaborations on related work.
This research was supported by the DFG through AN 275/5-1 (FBA),
SFB 484 (RB), SFB 608 (MV),
as well as the NIC, FZ J\"ulich under project no. HHB000 (FBA).




\begin{thebibliography}{}
\bibitem{entropyfoot}
Impurity entropy is defined as the impurity contribution to the
thermodynamic entropy of the model (\ref{eq:sbm}).
\bibitem{nrgpara}
The NRG parameters are:
$\Lambda\!=\!2$, $N_s\!=\!100$, $N_b\!=\!8$, $N_{b0}\!=\!100$.
The non-equilibrium calculations employ
$N_Z\!=\!32$ and no additional damping,
see Ref.~\protect\onlinecite{neqnrg} for details.
\end{thebibliography}
\end{document}